\documentclass[a4paper,12pt]{article}
\usepackage{authblk}

\usepackage{amssymb}
\usepackage{amsmath}

\usepackage{tikz}
\usetikzlibrary{math,decorations.pathreplacing,calligraphy}
\usepackage{diagbox}
\usepackage[width=16.00cm]{geometry}
\def\figwidth{14cm}

\def\refp#1{(\ref{#1})}
\def\showeps{false}
\def\half{\frac{1}{2}}
\def\new{\tilde}

\begin{document}

\title{Combined WENO schemes for increasing the accuracy of the numerical solution of conservation laws}

\author[1]{Hossein Mahmoodi Darian\thanks{hmahmoodi@ut.ac.ir}}

\affil[1]{School of Engineering Science, College of Engineering, University of Tehran}

\date{}

\maketitle

\begin{abstract}
In this article, we introduce a new method which allows utilizing all the available sub-stencils of a WENO scheme to increase the accuracy of the numerical solution of conservation laws while preserving the non-oscillatory property of the scheme. In this method, near a discontinuity, if there is a smooth sub-stencil with higher-order of accuracy, it is used in the reconstruction procedure. Furthermore, in smooth regions, all the sub-stencils of the same order of accuracy form the stencil with the highest order of accuracy as the conventional WENO scheme. The presented method is assessed using several test cases of the linear wave equation and one- and two-dimensional Euler's equations of gas dynamics. In addition to the original weights of WENO schemes, the WENO-Z approach is used. The results show that the new method increases the accuracy of the results while properly maintaining the ENO property.
\end{abstract}

\subsubsection*{keywords:}
WENO schemes, Shock-capturing schemes, Nonlinear weights

\section{Introduction}
\label{intro}
Numerical solution of hyperbolic systems of conservation laws due to presence of shocks and other discontinuities is a difficult task, especially when using high-order accurate schemes. Among various shock-capturing schemes, Weighted Essentially Non-Oscillatory (WENO) schemes are very popular. They were first introduced in \cite{liu,jiang-shu} and gradually different modifications and improvements over the original ones were proposed in the literature.
Henrik et al. \cite{HENRICK2005542} showed that near the critical points, the original weights are not close enough to their optimal values and cannot achieve the maximum accuracy.
They introduced mapped WENO schemes (WENO-M), where they used a mapping function for the weights of \cite{jiang-shu} to recover the maximum order in the critical points. This work made a basis for introducing new mapping functions \cite{Feng2012,FENG2014453,Wang2016540,US2019162}. Another improvement was the WENO-Z scheme, which was introduced by Borges et al. \cite{BORGES20083191} and then generalized by Castro et al. \cite{CASTRO20111766}. In contrast to \cite{HENRICK2005542} and its similar work, they defined new weights, instead of mapping the original weights of \cite{jiang-shu}.
Balsara et al. \cite{BALSARA2016780} made a non-linear hybridization between a large, centered, very high accuracy stencil and a lower order WENO scheme that is nevertheless very stable and capable of capturing physically meaningful extrema and called their method WENO-AO (Adaptive Order).
Recently, Amat et al. in a series of papers \cite{Amat20191205,Amat2020,Amat2020661,Amat2021,Amat2022}
introduced a new WENO interpolation capable of raising the order of accuracy close to discontinuities for some applications such as data interpolation and signal processing. Their idea was to use the highest-order stencil as long as the stencil does not contain any discontinuity. They used their new WENO algorithm for the numerical solution of conservation laws in \cite{Amat2021172}.

In this manuscript, we introduce a general method for using all the available stencils of different orders to increase the accuracy of WENO schemes while preserving the non-oscillatory property of the scheme. The idea is to first compute the flux using a weighted combination of all the sub-stencils of the same order of accuracy and then combine the resulted fluxes to form the final flux. We denote this scheme as combined WENO scheme (WENO-C).

The manuscript is organized as follows. In Section \ref{sec:1}, we give a brief description of the WENO schemes. The main part of this work will be presented in Section \ref{sec:new}, where we introduce a general formula for combining different stencils of different orders. In Section \ref{experiments}, the new scheme is assessed by numerical simulation of the linear advection equation and Euler equations of gas dynamics. Finally, the concluding remarks are given in Section \ref{conclusions}.

\section{WENO schemes}
\label{sec:1}
In this section, we describe the WENO schemes for discretizing the governing equation of a conservation law. Consider the following equation: 
\begin{eqnarray}
	\label{e201}
	u_t + f_x=0
\end{eqnarray} 
where $u$ is the conservative variable and $f=f(u)$ is the flux function. We consider a uniform grid in the $x$-direction and define the grid points as $x_i = i\Delta x$ where $\Delta x$ is the grid size. The flux derivative is discretized in the conservative form as 
\begin{eqnarray}
	\label{e:derv}
	f'(x_i) = \frac{f_{i+\half}-f_{i-\half}}{\Delta x} + O(\Delta x^m)
\end{eqnarray} 
where $f_{i+\half}$ is the numerical flux at $x_{i+\half}$ and $m$ is the order of the truncation error. In a $(2k-1)$th-order WENO scheme \cite{liu,jiang-shu}, the flux $f_{i+\half}$ is approximated such that $m=2k-1$ in smooth regions and $m=k$ near discontinuities. This is done by a convex combination of $k$ fluxes:
\begin{eqnarray}
	\label{e:comb}
	f_{i+\half} = \sum_{r=0}^{k-1} \omega_r f_{i+\half}^{(k,r)} \;, \qquad 0\leq r\leq k-1 
\end{eqnarray}
where $f_{i+\half}^{(k,r)}$ is a $k$th-order approximation of the flux. The flux $f_{i+\half}^{(k,r)}$ satisfies \refp{e:derv} with $m=k$ by using the following stencil points
\begin{eqnarray}
	\label{e:stencil}
	S^{k}_r = \left\{x_{i-r},x_{i-r+1},\cdots,x_{i-r+k-1}\right\}
\end{eqnarray}
and it is obtained by a polynomial of degree at most $k-1$, $p^k_r(x)$, which its average in $[x_{j-\half},x_{j+\half}]$ equals $f_j$ for all $x_j\in S^k_r$:
\begin{eqnarray}
	\label{e:poly}
	f_{j} = \frac{1}{\Delta x}\int_{x_{j-\half}}^{x_{j+\half}} p^k_r(\xi)\;d\xi \;,\qquad x_j\in S^k_r
\end{eqnarray}

The coefficients $\omega_r$ are nonlinear weights which determine the contribution of each $f_{i+\half}^{(k,r)}$ to $f_{i+\half}$. For consistency and stability, it is required:
\begin{eqnarray}
	\label{e:consistency}
	\omega_r \geq 0 \;,\qquad	\sum_{r=0}^{k-1} \omega_r =1 
\end{eqnarray}
 
In \cite{jiang-shu} after extensive numerical experiments, the following weights are proposed 
\begin{eqnarray}
	\label{e:weights}
	\omega_r = \frac{\alpha_r}{\sum\nolimits_{q=0}^k \alpha_q}\;, \qquad
	\alpha_r = \frac{d_r}{(\beta_r+\epsilon)^2} 
\end{eqnarray}
where the coefficients $d_r$ are optimal weights which generate the $(2k-1)$th-order central upwind scheme:
\begin{eqnarray}
	\label{e:combopt}
	f_{i+\half}^{(2k-1,k)} = \sum_{r=0}^{k-1} d_r f_{i+\half}^{(k,r)} 
\end{eqnarray}

The coefficients $\beta_r$ are called the smoothness indicators and defined by
\begin{eqnarray}
	\label{e:SI}
	\beta_r = \sum_{l=1}^{k-1}\int_{x_{i-\half}}^{x_{i+\half}}\Delta x^{2l-1}
	\left(\frac{\partial^l p^k_r(x)}{\partial x^l}\right)^2  dx 
\end{eqnarray}

In the literature, this method  is usually represented by WENO-JS in honor of the authors of \cite{jiang-shu}. Another method is the WENO-Z scheme \cite{BORGES20083191,CASTRO20111766} which differs from the WENO-JS scheme in the definition of $\alpha_r$:
\begin{eqnarray}
	\label{e:wenoz}
	\alpha_r &=& d_r \left(1+\left(\frac{\tau_{2k-1}}{\beta_r+\epsilon}\right)^2 \right)\\
	\label{e:tau}
	\tau_{2k-1} &=& \left\{
\renewcommand{\arraystretch}{1.75}
	\begin{array}{lr}
		|\beta_0 - \beta_{k-1}| & \mod(k,2)=1 \\	
		|\beta_0 - \beta_{1} - \beta_{k-2} + \beta_{k-1}| & \mod(k,2)=0 
	\end{array}
	\right.
\end{eqnarray}

\section{The new scheme}
\label{sec:new}
\begin{figure}\begin{center}		
	\includegraphics[width=\figwidth]{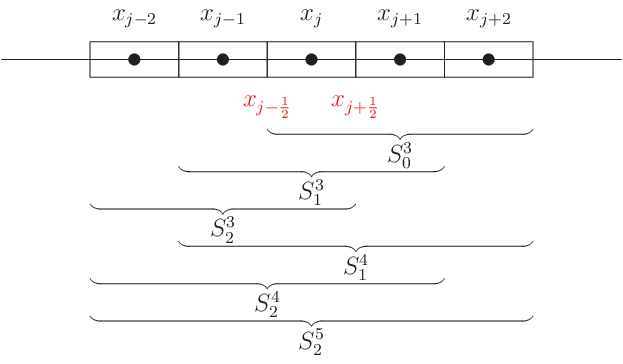}
	\caption{All sub-stencils inside $S^5_2$ ($k=3$).}
	\label{f:stencils:5}
\end{center}\end{figure}

\begin{figure}\begin{center}		
	\includegraphics[width=\figwidth]{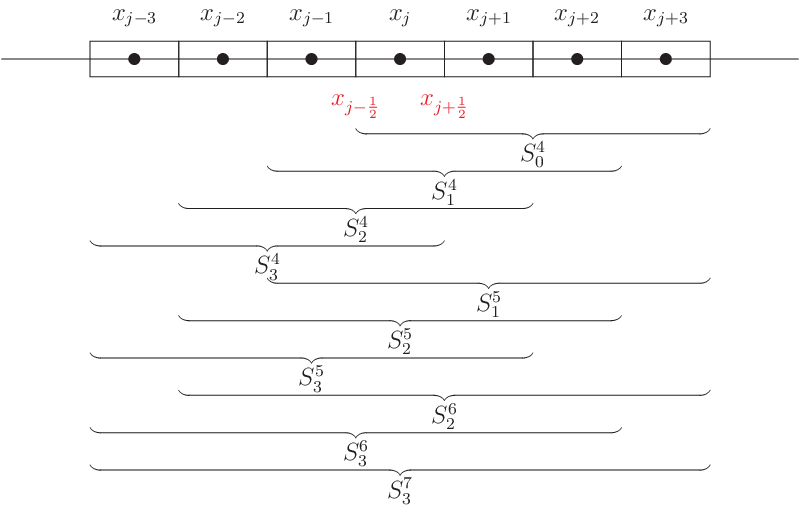}
	\caption{All sub-stencils inside $S^7_3$ ($k=4$).}
	\label{f:stencils:7}
\end{center}\end{figure}

As described in the previous section, in a $(2k-1)$th-order WENO scheme, a linear combination of $k$th-order sub-stencils ($S_r^k$) form the $(2k-1)$th-order flux \refp{e:combopt}. These sub-stencils are inside the stencil $S_{k-1}^{2k-1}$. However, there are other sub-stencils inside $S_{k-1}^{2k-1}$, which can form the same $(2k-1)$th-order flux. For instance, Figs.~\ref{f:stencils:5} and \ref{f:stencils:7} show the sub-stencils inside  $S_2^{5}$ and  $S_3^{7}$, respectively.

More precisely, there are $(k-s)$ sub-stencils of $(k+s)$th-order inside $S_{k-1}^{2k-1}$:
\begin{eqnarray}
	\label{e:stencils}
	\left\{S_{s}^{k+s},S_{s+1}^{k+s},\cdots,S_{k-1}^{k+s} \right\}\;, \qquad 0\leq s\leq k-1
\end{eqnarray}
which can form 
\begin{eqnarray}
	\label{e:comboptnew}
	f_{i+\half}^{(2k-1,k)} = \sum_{r=s}^{k-1} d^{(k+s)}_r f_{i+\half}^{(k+s,r)}  
\end{eqnarray}

Note that for $s=0$, \refp{e:comboptnew} reduces to \refp{e:combopt}. Leaving aside the largest stencil ($s=k-1$), we have $k-1$ linear combinations to construct $f_{i+\half}^{(2k-1,k)}$. For each $s$, we use the same procedure of the conventional WENO schemes in \refp{e:weights} to form a weighted linear combination of $(k+s)$th-order fluxes:
\begin{eqnarray}
	\label{e:fluxes:s}
	\new f_{i+\half}^{(k+s)} &=& \sum_{r=s}^{k-1} \omega^{(k+s)}_r f_{i+\half}^{(k+s,r)} \\
	\label{e:JS}
	\omega_r^{(k+s)} &=& \frac{\alpha_r^{(k+s)}}{\sum\nolimits_{q=0}^k \alpha_q^{(k+s)}} \; , \qquad
	\alpha_r^{(k+s)} = \frac{d_r^{(k+s)}}{(\beta_r^{(k+s)}+\epsilon)^2} 
\end{eqnarray}
where the coefficients $d_r^{(k+s)}$ are optimal weights which generate the $(2k-1)$th-order central upwind scheme $f^{(2k-1,k)}$. Table \ref{t:optimal} shows the optimal weights for $k=3$ and $k=4$ which corresponds to the fifth- and seventh-order schemes, respectively. Note that the values in rows $s=0$ are the known optimal weights of the WENO5 and WENO7 schemes. To distinguish between the  flux $\new f^{(k+s)}$ in \refp{e:fluxes:s} and the sub-stencil fluxes $f^{(k+s,r)}$, we call the former the intermediate flux. 

\begin{table}[h]
	\caption{Optimal weights for $k=3$ and $k=4$.} 
\label{t:optimal}
\centering
\renewcommand{\arraystretch}{1.5}
\begin{tabular}{|c|c|c|c|c|c|}\hline
	$k$ & $s$ & $d_0^{(k+s)}$ &$d_1^{(k+s)}$&$d_2^{(k+s)}$&$d_3^{(k+s)}$ \\\hline
	3& 0 & $\frac{3}{10}$ & $\frac{6}{10}$ & $\frac{1}{10}$ & ---\\\cline{2-6}
	& 1 & --- &$\frac{3}{5}$ & $\frac{2}{5}$ & --- \\\hline
	4&0 & $\frac{4}{35}$ & $\frac{18}{35}$ & $\frac{12}{35}$ & $\frac{1}{35}$\\\cline{2-6}
	&1 & --- & $\frac{2}{7}$ & $\frac{4}{7}$ & $\frac{1}{7}$  \\\cline{2-6}
	&2 & --- & ---& $\frac{4}{7}$ & $\frac{3}{7}$ \\\hline
\end{tabular}
\renewcommand{\arraystretch}{1}
\end{table}

It is also possible to use the WENO-Z approach \cite{CASTRO20111766} to define $\alpha_r^{(k+s)}$:
\begin{eqnarray}
	\label{e:wenoz:s}
	\alpha_r^{(k+s)} &=& d_r^{(k+s)} \left(1+\left(\frac{\tau_{2k-1}}{\beta_r^{(k+s)}+\epsilon}\right)^2 \right)\\
	\label{e:tau:s}
	\tau_{2k-1} &=& \left\{
\renewcommand{\arraystretch}{1.75}
	\begin{array}{lr}
		\left|\beta_0^{(k)} - \beta_{k-1}^{(k)}\right| & \mod(k,2)=1 \\
		\left|\beta_0^{(k)} - \beta_{1}^{(k)} - \beta_{k-2}^{(k)} + \beta_{k-1}^{(k)}\right| & \mod(k,2)=0 
	\end{array}
	\right.
\end{eqnarray}
where in the definition of $\tau_{2k-1}$, we use only the smoothness indicators of the $k$th-order sub-stencils. 

Now, we use the intermediate fluxes \refp{e:fluxes:s} to construct the final flux:
\begin{eqnarray}
	\label{e:newflux}
	f_{i+\half} = \sum_{s=0}^{k-2} \gamma^{(k+s)} \new f_{i+\half}^{(k+s)} 
\end{eqnarray}
where $\gamma^{(k+s)}$ are the weights which to be determined. We call these weights the total weights to distinguish them from the weights $\omega_r^{(k+s)}$.  Furthermore, we denote this method by WENO-C where C is the first letter of ``Combined''. Also, if instead of \refp{e:JS}, the WENO-Z approach \refp{e:wenoz:s} is used  for computing the coefficients $\alpha_r^{(k+s)}$, then we denote the new scheme by WENO-ZC.

To have proper total weights ($\gamma^{(k+s)}$) for each of the intermediate fluxes in \refp{e:newflux}, we use the same weighted linear combination as \refp{e:fluxes:s} to define the corresponding smoothness indicators:
\begin{eqnarray}
	\new\beta^{(k+s)} = \sum_{r=s}^{k-1} \omega^{(k+s)}_r \beta_r^{(k+s)} 
\end{eqnarray}

This definition (for a fixed $s$) relates the smoothness indicator $\new\beta^{(k+s)}$ to the smoothness indicators of its sub-stencils. We call $\new\beta^{(k+s)}$ the total smoothness indicator of the $(k+s)$th-order sub-stencils. Therefore, the smoothness contribution of each sub-stencil is equal to the contribution of its corresponding flux to the intermediate flux \refp{e:fluxes:s}. This means, when comparing the stencils of two different intermediate fluxes (different $s$), the comparison is mostly made between the sub-stencils which have the greatest influence on that intermediate flux. In other words, if a sub-stencil flux ($f^{(k+s,r)}$) is ruled out due to intersecting a discontinuity, then its smoothness contribution is also eliminated from $\new\beta^{(k+s)}$. 

Now, we define the weights $\gamma^{(k+s)}$ in a similar approach of the original WENO schemes as:
\begin{eqnarray}
	\gamma^{(k+s)} = \frac{\new\alpha^{(k+s)}}{\sum\nolimits_{q=0}^{k-2} \new\alpha^{(k+q)}} \; , \qquad
	\new\alpha^{(k+s)} = \frac{\new d^{(k+s)}}{(\new\beta^{(k+s)}+\epsilon)^2} 
\end{eqnarray}
where the coefficients $\new d^{(k+s)}$ determine the preference of each $\new f^{(k+s)}$ in smooth regions. Since, in smooth regions, all the intermediate fluxes in \refp{e:newflux} reduce to the $(2k-1)$th-order accurate flux $f^{(2k-1,k)}$, there is no preference between them. 
However, near discontinuities, it is desirable for small deviationof the weights from their optimal values in \refp{e:fluxes:s}, the intermediate flux with higher-order sub-stencils is assigned a larger total weight as long as it does not spoil the ENO property of the method. Through some numerical experiments, we propose 
\begin{eqnarray}
	\new d^{(k+s)} = (1+s)^p \;, \qquad 0\leq s\leq k-2
\end{eqnarray}
where $p$ is a positive number. Therefore, it is an increasing function of $s$. Also, larger values of $p$ assign larger values to the total weight of the intermediate fluxes with higher-order sub-stencils (larger $s$).

\section{Numerical experiments}
\label{experiments}
In this section, we assess the numerical performance of the new WENO scheme. We set $\epsilon=10^{-12}$ for WENO-JS and WENO-C and $\epsilon=10^{-40}$ for WENO-Z and WENO-ZC. 
For the time integration, the following third-order TVD Runge-Kutta scheme \cite{shu-osher-1} is used 
\begin{eqnarray}
	u^{(1)} &=& u^n + \Delta t L(u^{n}) \nonumber \\
	u^{(2)} &=& \frac{3}{4}u^n+\frac{1}{4}u^{(1)}+ \frac{1}{4}\Delta t L(u^{(1)}) \nonumber \\
	u^{n+1} &=& \frac{1}{3}u^n+\frac{2}{3}u^{(2)}+ \frac{2}{3}\Delta t L(u^{(2)}) 
	\label{e103}
\end{eqnarray} 
where $\Delta t$ is the time-step.

\subsection{Advection equation}\label{s.wave}
The first test case is the linear advection equation:
\begin{eqnarray}
	u_t+u_x=0, \qquad x\in[-1,1]
\end{eqnarray}

The domain is periodic and the initial condition is a function which contains a Gaussian-, a triangle-, a square- and a half-ellipse-wave region \cite{jiang-shu}:
\begin{eqnarray}
	\renewcommand{\arraystretch}{1.5}
	u(x,0) = {\left\{
		\begin{array}{ll}
			\frac{1}{6}(Q(x,\beta,\zeta-\delta)+Q(x,\beta,\zeta+\delta)+4Q(x,\beta,\zeta)), &  x\in[-0.8, -0.6] \\
			1, &  x\in [-0.4, -0.2] \\
			1-|10(x-0.1)|, & \,~~x\in[0.0,0.2] \\
			\frac{1}{6}(R(x,\alpha,a-\delta)+R(x,\alpha,a+\delta)+4R(x,\alpha,a)), & \,~~x\in [0.4,0.6] \\
			0, & \,~~{\rm otherwise}
		\end{array}\right.} \nonumber\\
	\label{e:wave:init}
\end{eqnarray}
where
\begin{eqnarray}
Q(x,\beta,\zeta) = \exp(-\beta(x-\zeta)^2), \nonumber\\
R(x,\alpha,a) = \sqrt{\max(1-\alpha^2(x-a)^2,0)} \nonumber
\end{eqnarray}
and the constants are
\begin{eqnarray*}
	a=0.5\;, \quad \zeta=-0.7\;,\quad \delta=0.005\;,\quad \alpha=10\;,\quad \beta=\ln 2/(36\delta^2)
\end{eqnarray*}

\begin{figure}[!htbp]
	\centering
	\setkeys{Gin}{draft=\showeps}
	\includegraphics[width=\figwidth]{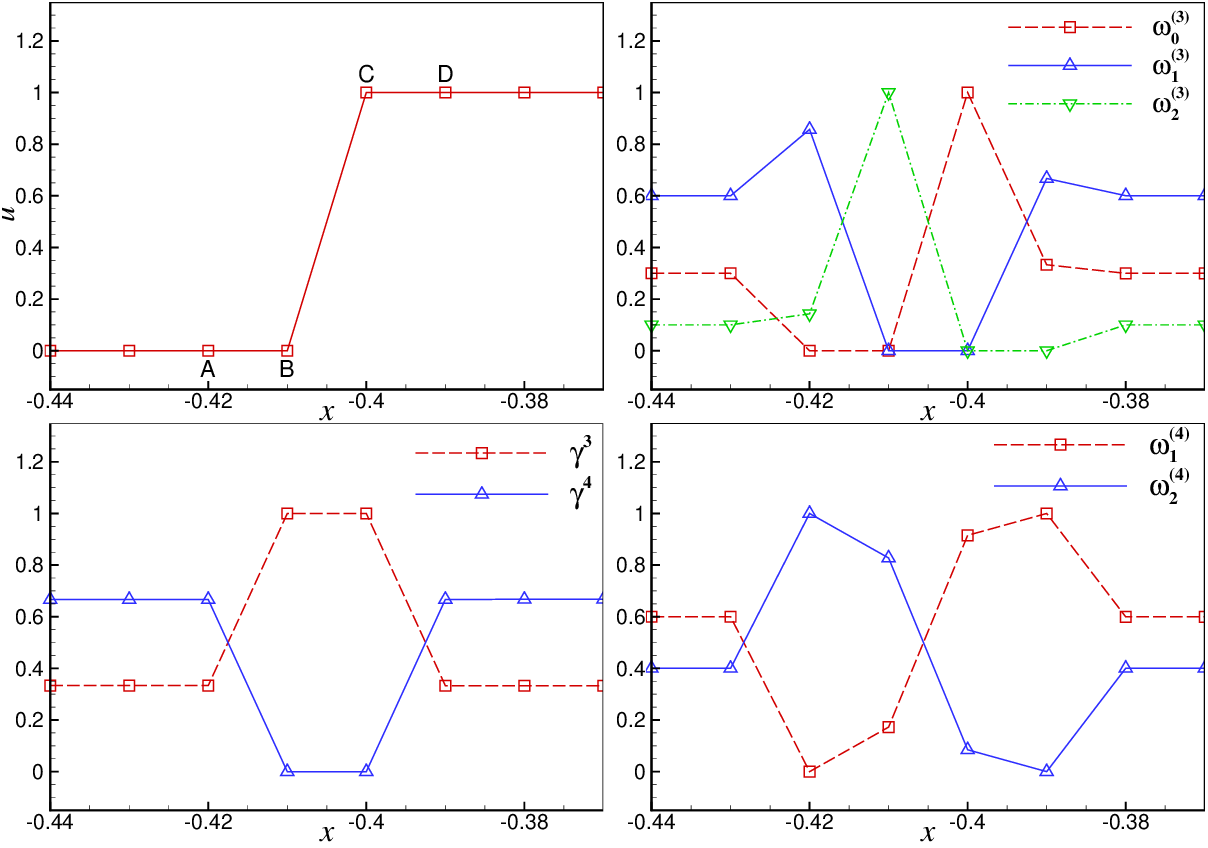}
	\caption{The distribution of the weights near a discontinuity;
		top-left) the function \refp{e:wave:init}, bottom-left) total weights of the third- and fourth-order stencils, top-right) weights of the third-order sub-stencils, bottom-right) weights of the fourth-order sub-stencils.}
	\label{f:weight:zoom} 
	\setkeys{Gin}{draft=\showeps}
	\includegraphics[width=\figwidth]{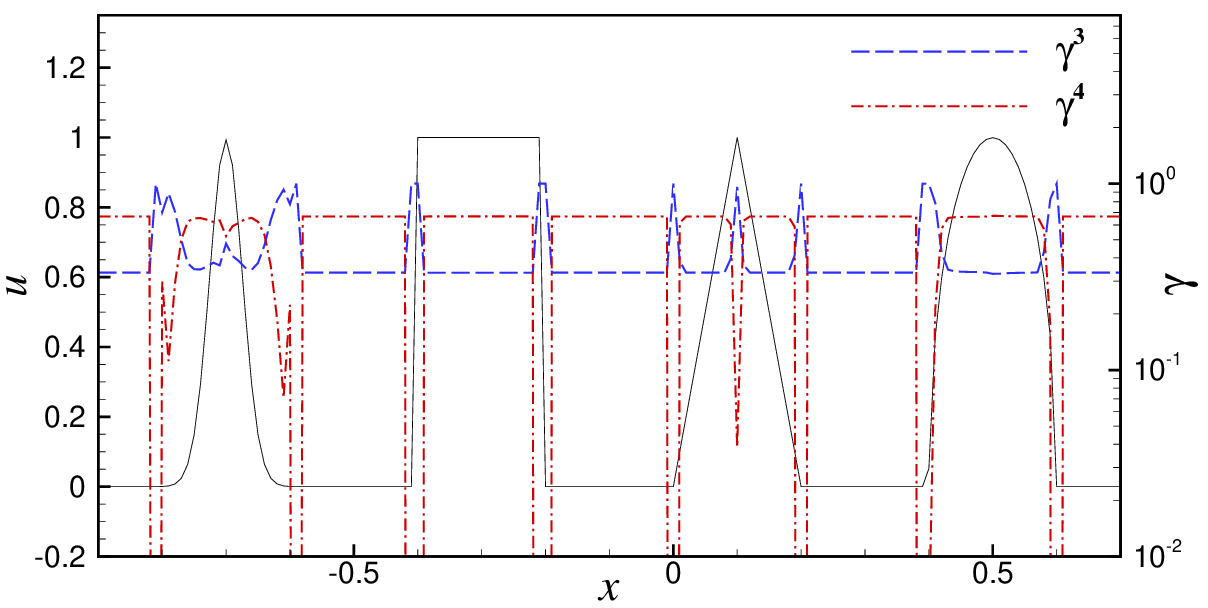}
	\caption{The distribution of the total weights $\gamma^3$ and $\gamma^4$ for the function \refp{e:wave:init}.}
	\label{f:weight} 
\end{figure}

First, we give a description on the weights. 
Figure \ref{f:weight:zoom} shows the distribution of the weights around the discontinuity of the square wave for the WENO5-C scheme with $p=1$. The points B and C at $x=-0.41$ and $x=-0.40$ are immediately before and after the discontinuity, respectively. At these points we have $\gamma^3 = 1$ and $\gamma^4 = 0$. This is because both the fourth-order sub-stencils (corresponding to these points) intersect the discontinuity. Therefore, the fourth-order sub-stencils are not used in the flux computation. Furthermore, the weights of the third-order sub-stencils for the point B are $(\omega^{(3)}_0,\omega^{(3)}_1,\omega^{(3)}_2)=(0,0,1)$ which means only $S^3_2$ is used to compute the flux (see Fig. \ref{f:stencils:5}). Similarly, this is the case, in the reverse order, for the point C. 
For the points A and D, respectively at $x=-0.42$ and $x=-0.39$, which are one point away from the discontinuity, we have $\gamma^3 = \new d^3 = \frac{1}{3}$ and $\gamma^4 = \new d^4 = \frac{2}{3}$ which means the fourth-order sub-stencils are taken into account for the flux computation. Now, we observe the weights of the fourth-order sub-stencils corresponding to the point A are $(\omega^{(4)}_1,\omega^{(4)}_2)=(0,1)$ which means only $S^4_2$ is used to compute the flux and $S^4_1$ is abandoned due to intersecting the discontinuity.

Figure \ref{f:weight} shows the overall distribution of the total weights $\gamma^3$ and $\gamma^4$ for the function \refp{e:wave:init}. As it is observed, the total weight of the fourth-order schemes $\gamma^4$ is always lower than $\new d^4 = \frac{2}{3}$. Therefore, to increase the usage of the fourth-order schemes, it is required to increase $\new d^4$, or $p$ in \refp{e:comboptnew}. As stated previously, in smooth regions, the value of the total weights are not important, because both the combination of the third- and fourth-order subs-stencils produce the fifth-order scheme. However, when the weights $\omega^{(3)}_r$ and $\omega^{(4)}_r$ deviates from their optimal values, it is desirable to use the fourth-order sub-stencils as long as it does not spoil the ENO property of the method. Therefore, in the numerical test cases, we examine several values for $p$.

\begin{figure}[!htbp]
	\centering
	\setkeys{Gin}{draft=\showeps}
	\includegraphics[width=\figwidth]{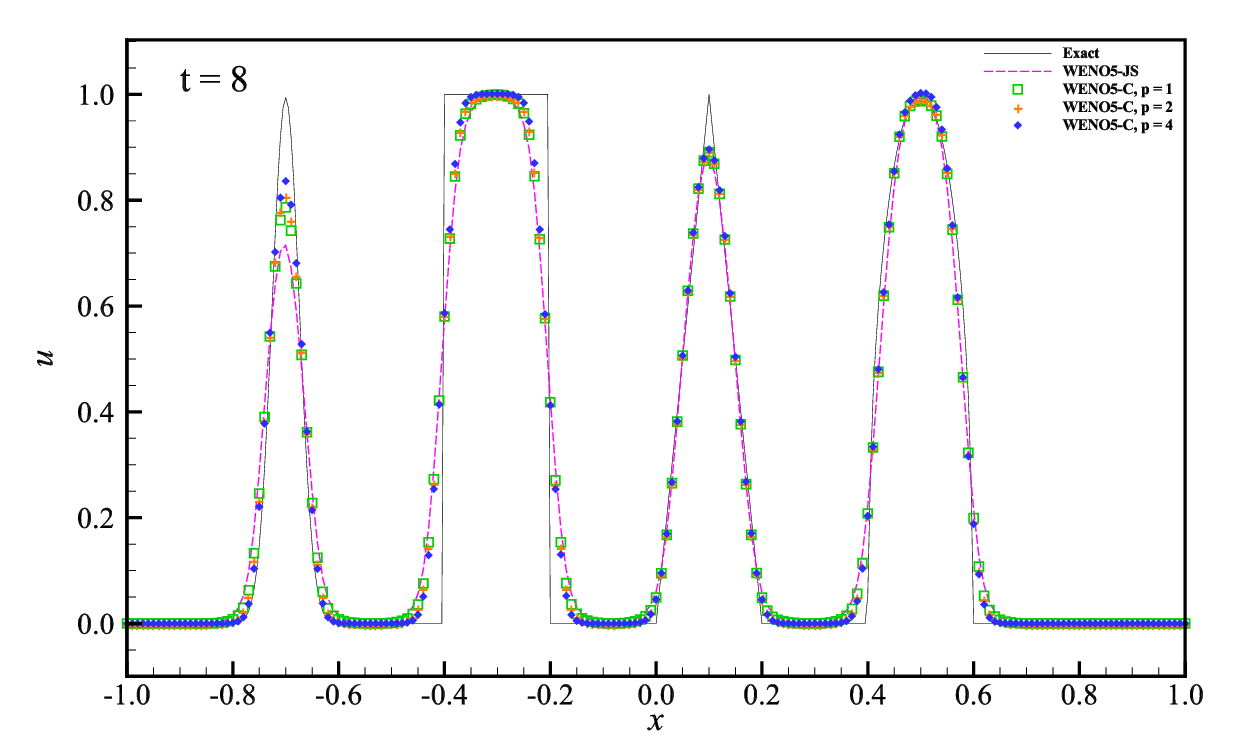}
	\caption{Numerical solution of the advection equation by the WENO5-JS and WENO5-C schemes using $N = 201$ at $t = 8$ using different values for the power parameter $p$.}
	\label{f:wave:8:5:JS} 
	\setkeys{Gin}{draft=\showeps}
	\includegraphics[width=\figwidth]{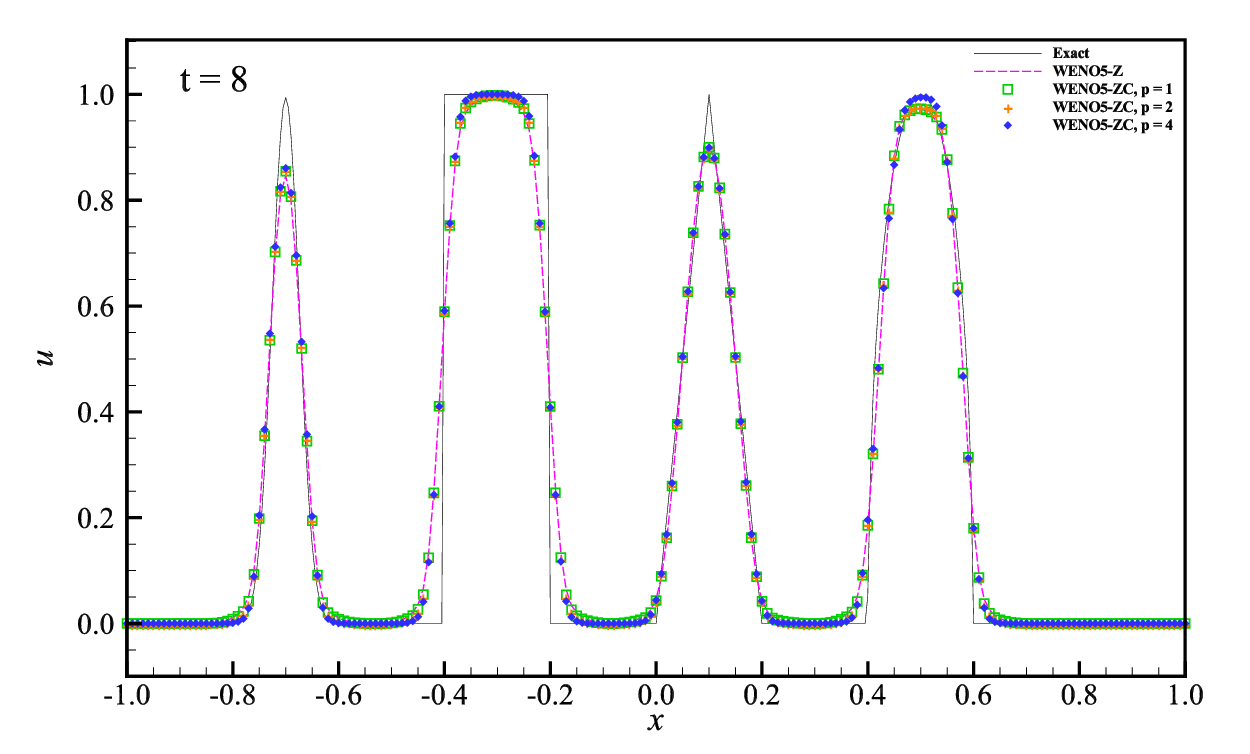}
	\caption{Numerical solution of the advection equation by the WENO5-Z and WENO5-ZC schemes using $N = 201$ at $t = 8$ using different values for the power parameter $p$.}
	\label{f:wave:8:5:Z} 
\end{figure}

\begin{figure}[!htbp]
	\centering
	\setkeys{Gin}{draft=\showeps}
	\includegraphics[width=\figwidth]{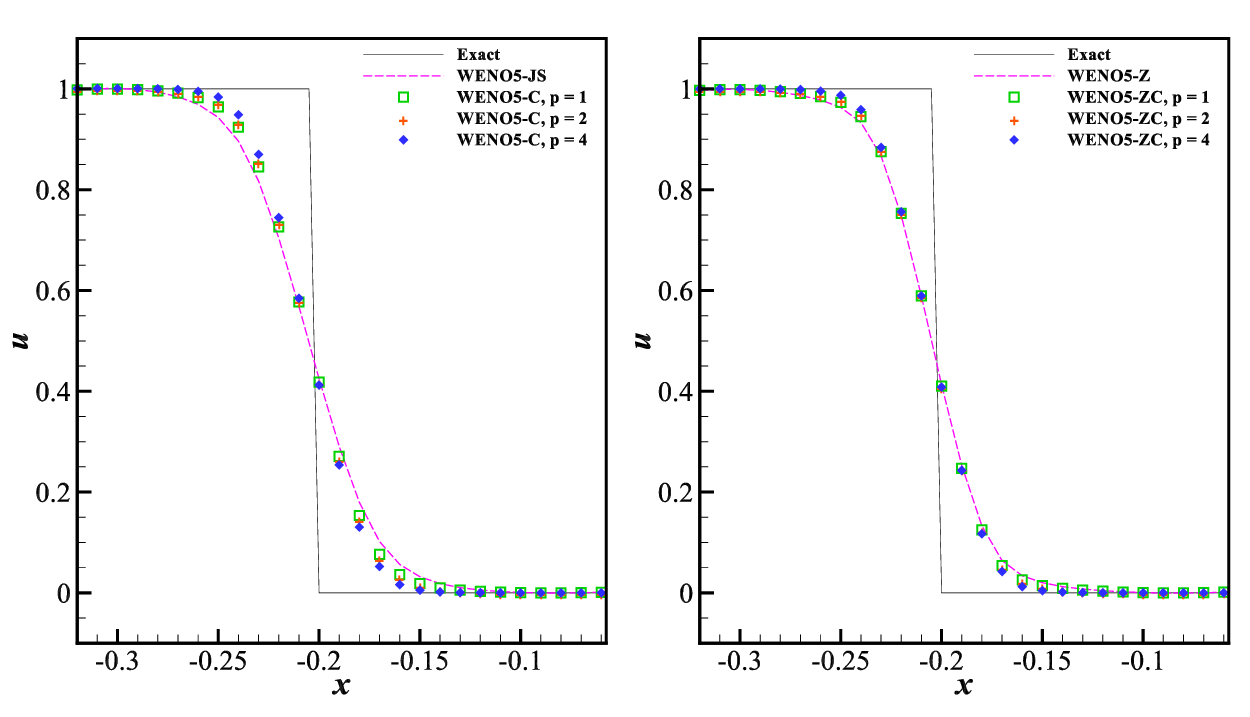}
	\caption{The zoomed view of Figs. \ref{f:wave:8:5:JS} (left) and \ref{f:wave:8:5:Z} (right) around the right discontinuity of the square wave.}
	\label{f:wave:8:5:JSZ} 
\end{figure}

Figure \ref{f:wave:8:5:JS} shows the numerical results at $t=8$ using $N=201$ for the WENO5 schemes along with their combined versions, WENO5-C, and different values for the power parameter $p$. Also, figure \ref{f:wave:8:5:Z} shows the same comparison for the WENO5-Z and WENO5-ZC schemes and a zoomed view of these figures around the right discontinuity of the square wave, are given in Fig.~\ref{f:wave:8:5:JSZ}. Figure \ref{f:wave:8:5:JS} shows the overall accuracy of the results increases by increasing $p$ and the ENO property of the schemes are preserved. This can be observed especially near the discontinuities of the square wave and also at the foot of the Gaussian, triangle and half-ellipse waves. Furthermore, the improvements are observed at the peak of these waves. Similarly, in Fig.~\ref{f:wave:8:5:Z}, we observe the improvement of the results for the WENO5-ZC schemes. However, since the WENO5-Z scheme is more accurate than the WENO5-JS scheme, this improvement is less than that of the WENO5-C schemes. However, it is worth mentioning that at the peak of the half-ellipse wave, we see significant improvement for the WENO5-ZC scheme for $p=4$.

\subsection{Shock-tube problem}
\label{s.tube}
The second test case is the Sod shock-tube problem~\cite{sod}. The governing equations are the one-dimensional Euler equations of gas dynamics: 
\begin{eqnarray}
	\label{e:euler}
	&&U_t+F_x=0, \quad
	U = \left( \begin{array}{c} \rho \\ \rho u \\ E \end{array} \right), \quad
	F = \left( \begin{array}{c} \rho u \\ \rho u^2+p \\ (E+p)u \end{array} \right) \\
	&&E = \rho (e + \frac{u^2}{2}), \quad p = \rho e (\gamma-1),\quad \gamma=1.4 \nonumber
\end{eqnarray}
where $u$, $\rho$, $p$ and $e$ denote the velocity, density, pressure and internal energy per unit mass, respectively. The solution domain is $[0,10]$ and the left and right states of the discontinuity at $t=0$ are
\begin{eqnarray}
	\label{e:sod:init}
	(\rho_L,u_L,p_L) = &(1,0,1) ,& x \leq 5 \\
	(\rho_R,u_R,p_R) = &(0.125,0,0.1) ,& x > 5 \nonumber
\end{eqnarray}
where the exact solution possesses a self-similar solution which consists of a shock, a contact discontinuity and an expansion fan.  

\begin{figure}[!htbp]
	\centering
	\setkeys{Gin}{draft=\showeps}
	\includegraphics[width=\figwidth]{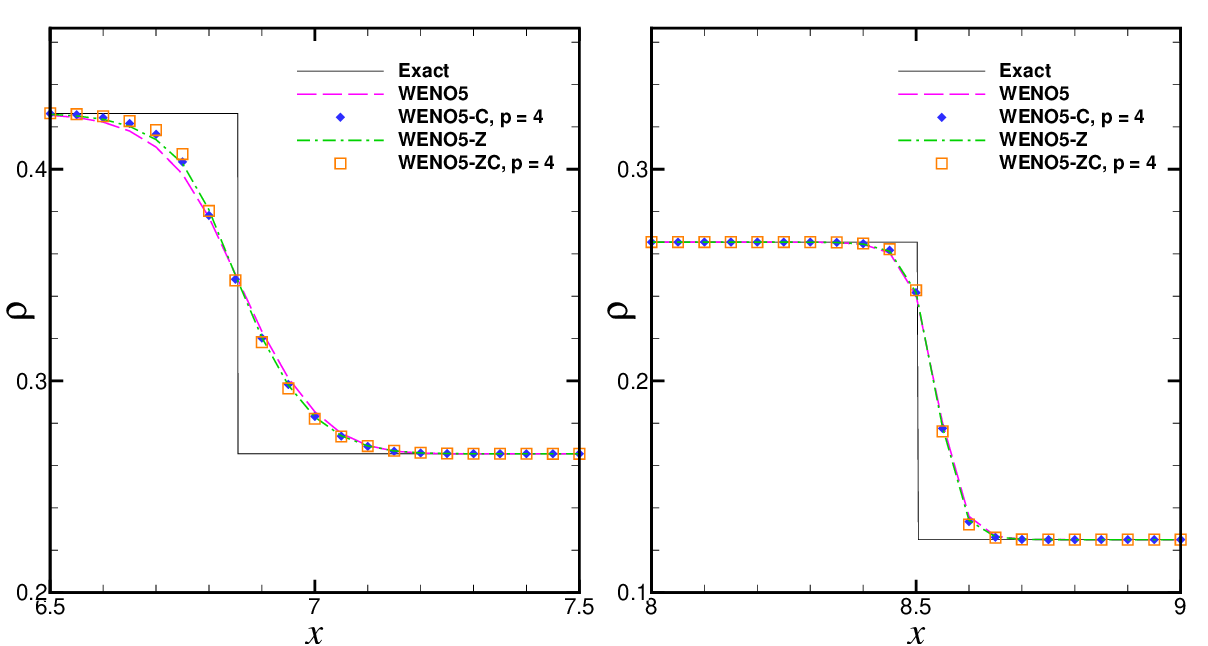}
	\includegraphics[width=\figwidth]{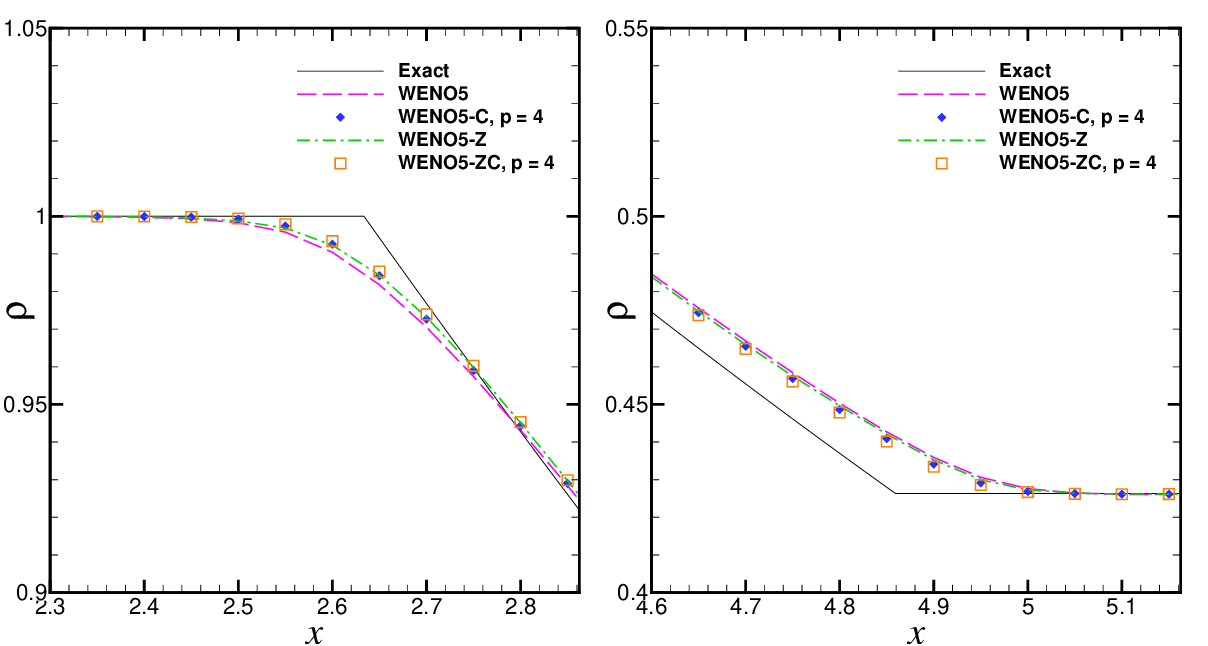}
	\caption{The density profile of the Sod shock-tube problem around the contact discontinuity (top-left), the shock (top-right) and the left and right ends of the expansion fan (bottom) using $N = 201$ at $t = 2$.}
	\label{f:sod:2:200} 
\end{figure}

Figure \ref{f:sod:2:200} shows the density profile at $t=2$ for different fifth-order schemes. The grid spacing is $\Delta x = 0.05$ ($201$ points) and the time marching is done using a fixed time-step of
$\Delta t = 0.01$. The characteristic-wise Lax-Friedrichs flux splitting \cite{weno-report} is used to handle both the negative and positive wave speeds. The results show the combined schemes (WENO5-C and WENO5-ZC) have more accurate results especially around the contact discontinuity ($x\simeq 6.7$) and the left and right ends of the expansion fan. It should be mention that increasing the accuracy is achieved without appearing any spurious oscillations.

\subsection{Shock-density wave interaction}
\label{s.density}
This test case involves the interaction between a Mach 3 moving shock and a density wave in shape of a sine function \cite{shu-osher-2}. The governing equations are the Euler equations (\ref{e:euler}). The solution domain is $[-5,5]$ and the initial states of the gas are as follows
\begin{eqnarray}
	\label{e:entropy:init}
	(\rho_L,u_L,p_L) = &(3.857143,2.629369,10.33333) ,& x \leq -4 \\
	(\rho_R,u_R,p_R) = &(1 + 0.2 \sin (5x),0,1) ,& x > -4 \nonumber
\end{eqnarray}

\begin{figure}[!htbp]
	\centering
	\setkeys{Gin}{draft=\showeps}
	\includegraphics[width=0.5\textwidth]{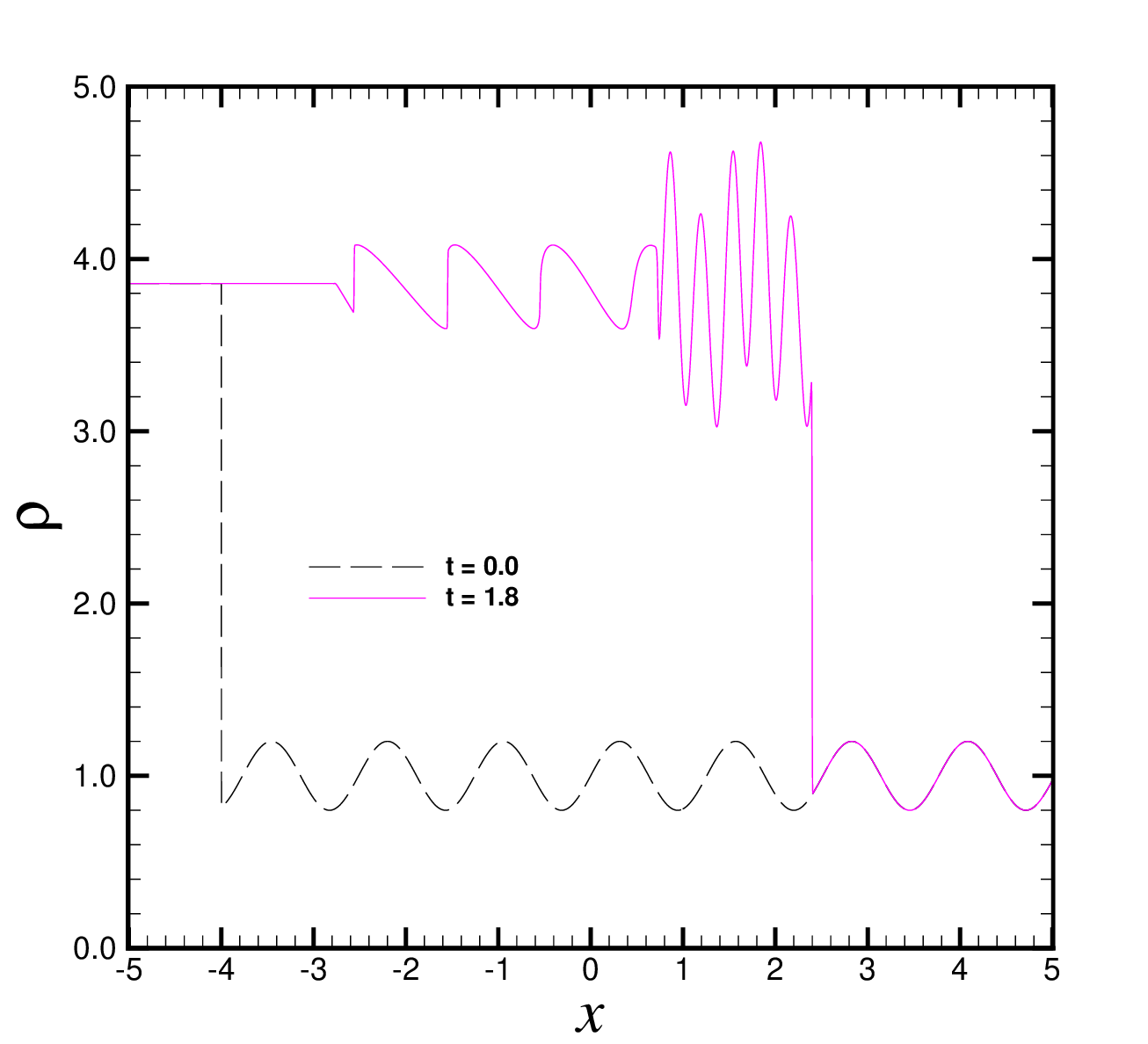}
	\caption{The density profile of the shock-density wave interaction problem at $t = 0$ and $t=1.8$.}
	\label{f:entropy:init}
\end{figure}

Figure~\ref{f:entropy:init} shows the density profile at the initial time and also at $t=1.8$. During the interaction, several shocks and high gradient regions appear. Specifically, at the time shown in the figure, the location of the moving shock is $x\approx 2.4$ and some shocks and steepening gradients are formed behind in $x\in[-2.9 , 0.8]$. Also, during  the interaction, the steepening gradients gradually form shocks. At the time $t=1.8$, two of these steepening gradients at $x\approx -2.6$ and $x\approx -1.6$, already become discontinuous. 

\begin{figure}[!htbp]
	\centering
	\setkeys{Gin}{draft=\showeps}
	\includegraphics[width=\figwidth]{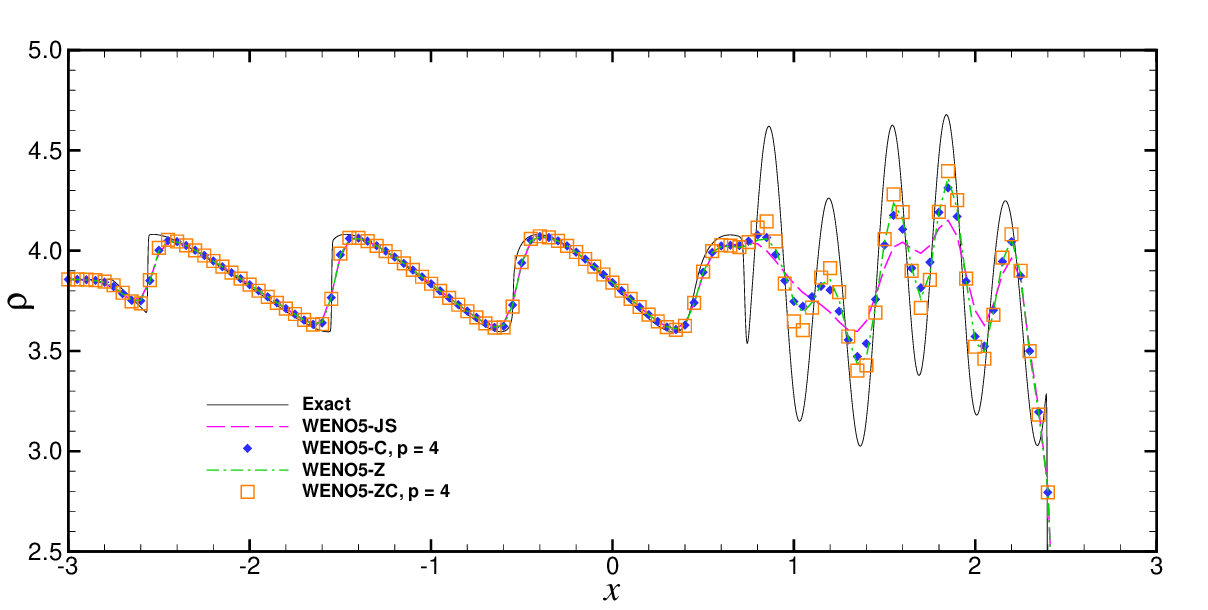}
	\caption{The density profile of the shock-density wave interaction problem using $N = 201$ at $t = 1.8$ for the fifth-order WENO schemes.}
	\label{f:entropy:2:200:5} 
\end{figure}

Figure \ref{f:entropy:2:200:5} shows the density distribution using $N=201$ points ($\Delta x = 0.05$) at $t=1.8$ for the fifth-order schemes. Also, the time-step is $\Delta t =0.002$. Comparing the results, especially in the high-gradient region, shows that WENO5-C and WENO5-ZC give considerably more accurate results than WENO5-JS and WENO5-Z, respectively. Furthermore, we observe the results of WENO5-ZC are more accurate that those of WENO5-C.

\begin{figure}[!htbp]
	\centering
	\setkeys{Gin}{draft=\showeps}
	\includegraphics[width=\figwidth]{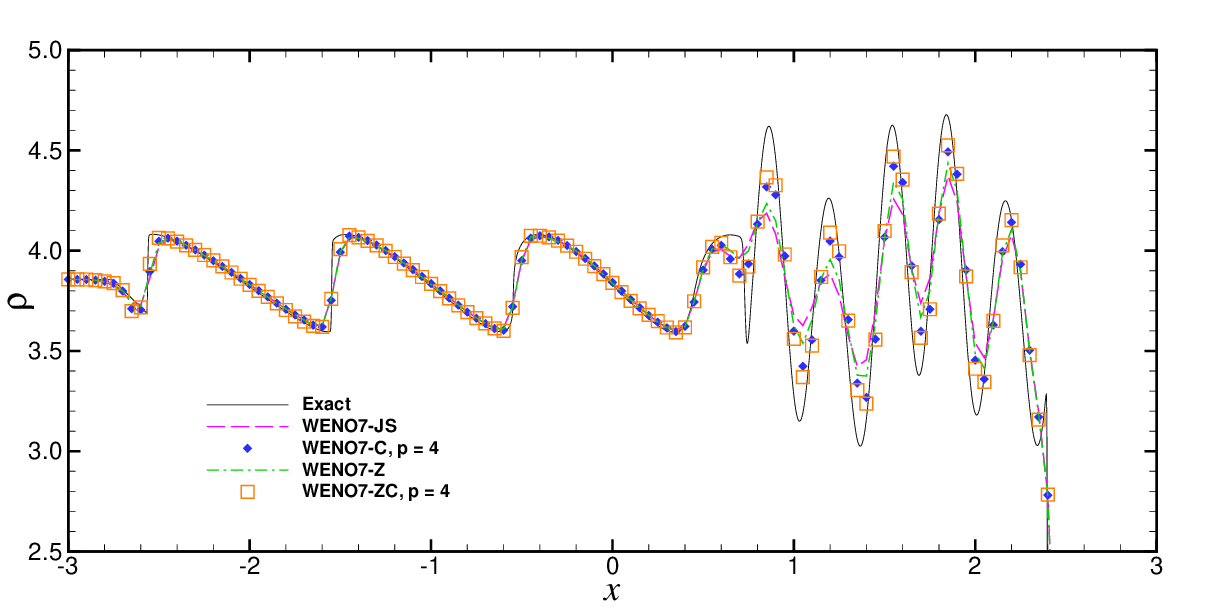}
	\caption{The density profile of the shock-density wave interaction problem using $N = 201$ at $t = 1.8$ for the seventh-order WENO schemes.}
	\label{f:entropy:2:200:7} 
\end{figure}

For this test case we also present the results for the seventh-order schemes in Fig.~\ref{f:entropy:2:200:7}. Again, we observe that the WENO7-C and WENO7-ZC schemes are less dissipative and give more accurate results in comparison with the WENO7-JS and WENO7-Z schemes, respectively. Also, the results of WENO7-ZC are more accurate than those of WENO7-C.

\subsection{Interacting blast waves}
\label{s.blast}
This is a very difficult test case for shock capturing schemes \cite{woodward}. It involves several interactions between strong shocks, rarefaction waves and contact discontinuities. The solution domain is $[0,1]$ and the initial condition consists of three constant states where the pressure in the middle part is significantly smaller than those in the left and right parts:
\begin{eqnarray}
	\label{e:blast:init}
	(\rho,u,p) = \left\{
	\begin{array}{lr}
		(1,0,1000) & 0.0\leq x <0.1 \\
		(1,0,0.01) & 0.1\leq x <0.9 \\
		(1,0,100) & 0.9\leq x \leq 1.0
	\end{array}
	\right.
\end{eqnarray}

The low pressure of the middle part, in case of appearing oscillations around the shocks, can cause negative pressure and therefore the simulation failure.

\begin{figure}[!htbp]
	\centering
	\setkeys{Gin}{draft=\showeps}
	\includegraphics[width=\figwidth]{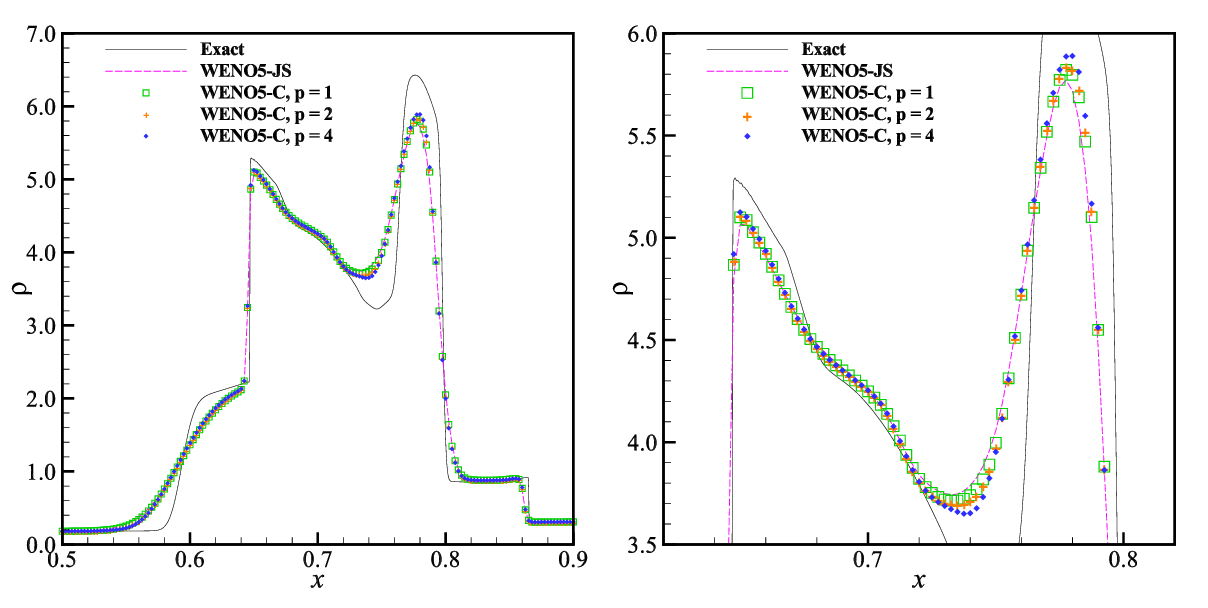}
	\caption{The density profile of the interacting blast waves problem using $N = 401$ at $t = 0.038$ for the fifth-order WENO schemes.}
	\label{f:blast:400:5} 
\end{figure}

Figure \ref{f:blast:400:5} compares the results of the fifth-order schemes at $t = 0.038$. The results are presented for $401$ grid points and $800$ time-steps with $\Delta t = 4.75\times 10^{-5}$. The results show that the WENO5-C schemes are more accurate than the WENO5-JS scheme especially at the peak and valley ($x \approx 0.74$ and $x \approx 0.78$ in the zoomed view). Also, close examination reveals a slightly sharper shock (at $x\approx 0.64$) is obtained using the WENO5-C scheme. Again, it is observed that the accuracy of the results is increased and the ENO property of the method is preserved. 

\subsection{Shock-vortex interaction}\label{s.vortex}
This is a two-dimensional test case where a vortex passes through a stationary normal shock~\cite{jiang-shu}. 
The governing equations are the two-dimensional Euler equations:
\begin{eqnarray}
	\label{e14}
	&&U_t+F_x+G_y=0, \quad \\
	&&U = \left( \begin{array}{c} \rho \\ \rho u \\ \rho v \\E \end{array} \right), \quad 
	F = \left( \begin{array}{c} \rho u \\ \rho u^2+p \\ \rho u v \\ (E+p)u \end{array} \right), \quad 
	G = \left( \begin{array}{c} \rho v \\ \rho v u \\ \rho v^2+p \\ (E+p)v \end{array} \right) \nonumber\\
	&&E = \rho (e + \frac{u^2+v^2}{2}), \quad p = \rho e (\gamma-1),\quad \gamma=1.4 \nonumber
\end{eqnarray}

The problem domain is $[0,2]\times [0,1]$ and the vertical shock is stationary and located at $x=0.5$. The shock Mach number is $1.1$. The left state of the shock is 
\begin{eqnarray}
	(\rho_L, u_L, v_L, p_L)=(1, 1.1\sqrt{\gamma}, 0, 1) \nonumber
\end{eqnarray}
and the right state is obtained using the normal-shock relations. A small isentropic vortex is superposed to the shock left state. The vortex center is at $(x_c, y_c)=(0.25, 0.5)$ and its properties (denoted by $\delta$) is added to the velocity $(u, v)$, temperature $(T = p/\rho)$, and entropy $(s = \ln(p/\rho^\gamma))$ of the shock left state:
\begin{eqnarray}
	(\delta u, \delta v) = \kappa\eta e^{\alpha(1-\eta^2)}(\sin\theta,-\cos\theta)\;,
	\quad \delta T = -\frac{(\gamma-1)\kappa^2e^{2\alpha(1-\eta^2)}}{4\alpha\gamma}\;,
	\quad \delta s = 0 \nonumber
\end{eqnarray}
where $\kappa=0.3$ is the vortex strength and $\eta=r/r_c$, where $r$ and $\theta$ are the polar coordinates with respect to the vortex center. Also, $r_c=0.05$ is the vortex critical radius and $\alpha=0.204$ is the vortex decay rate.

\begin{figure}[!htbp]
	\centering	
	\setkeys{Gin}{draft=\showeps}
	\includegraphics[width=0.46\textwidth]{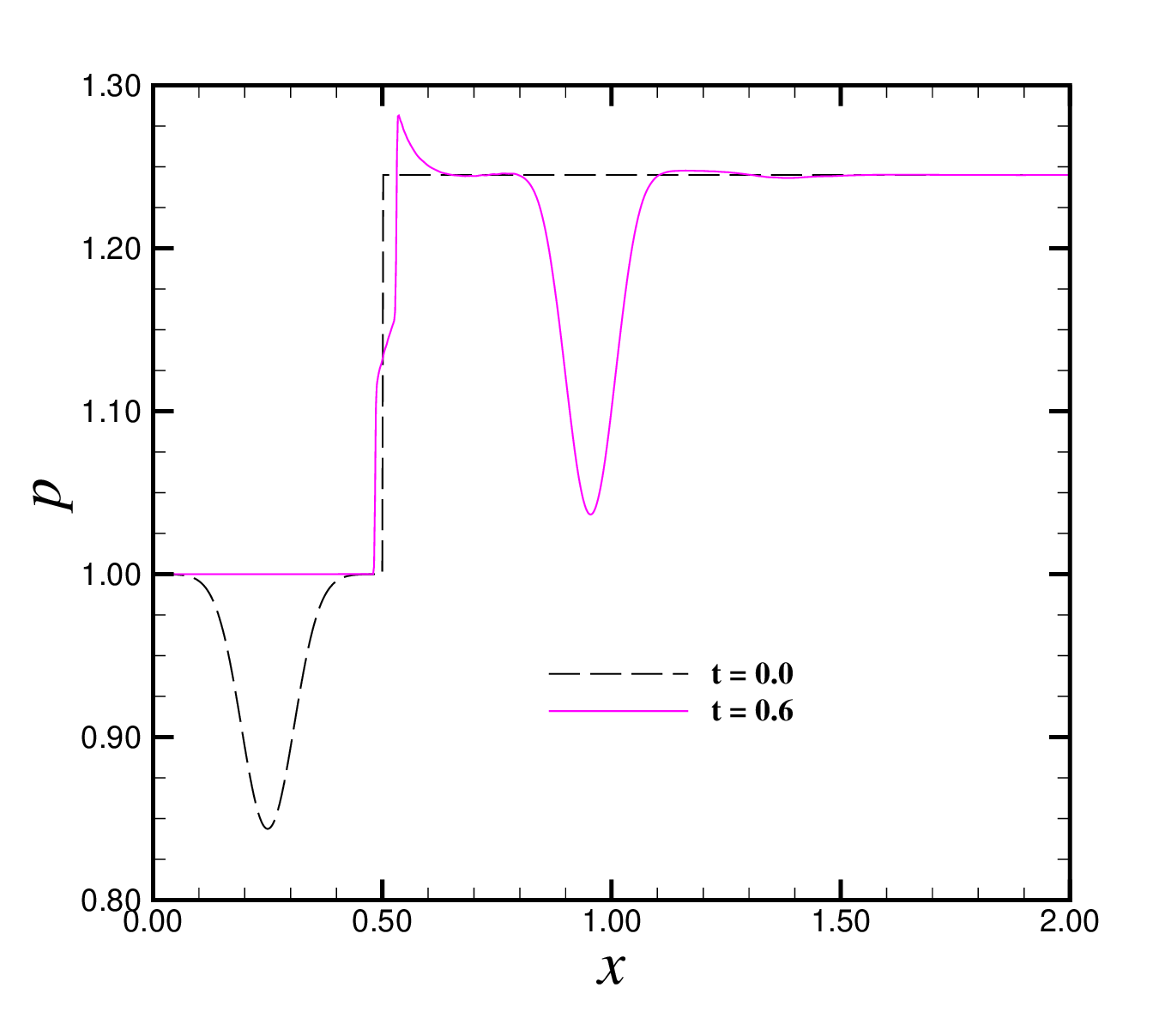}
	\includegraphics[width=0.46\textwidth]{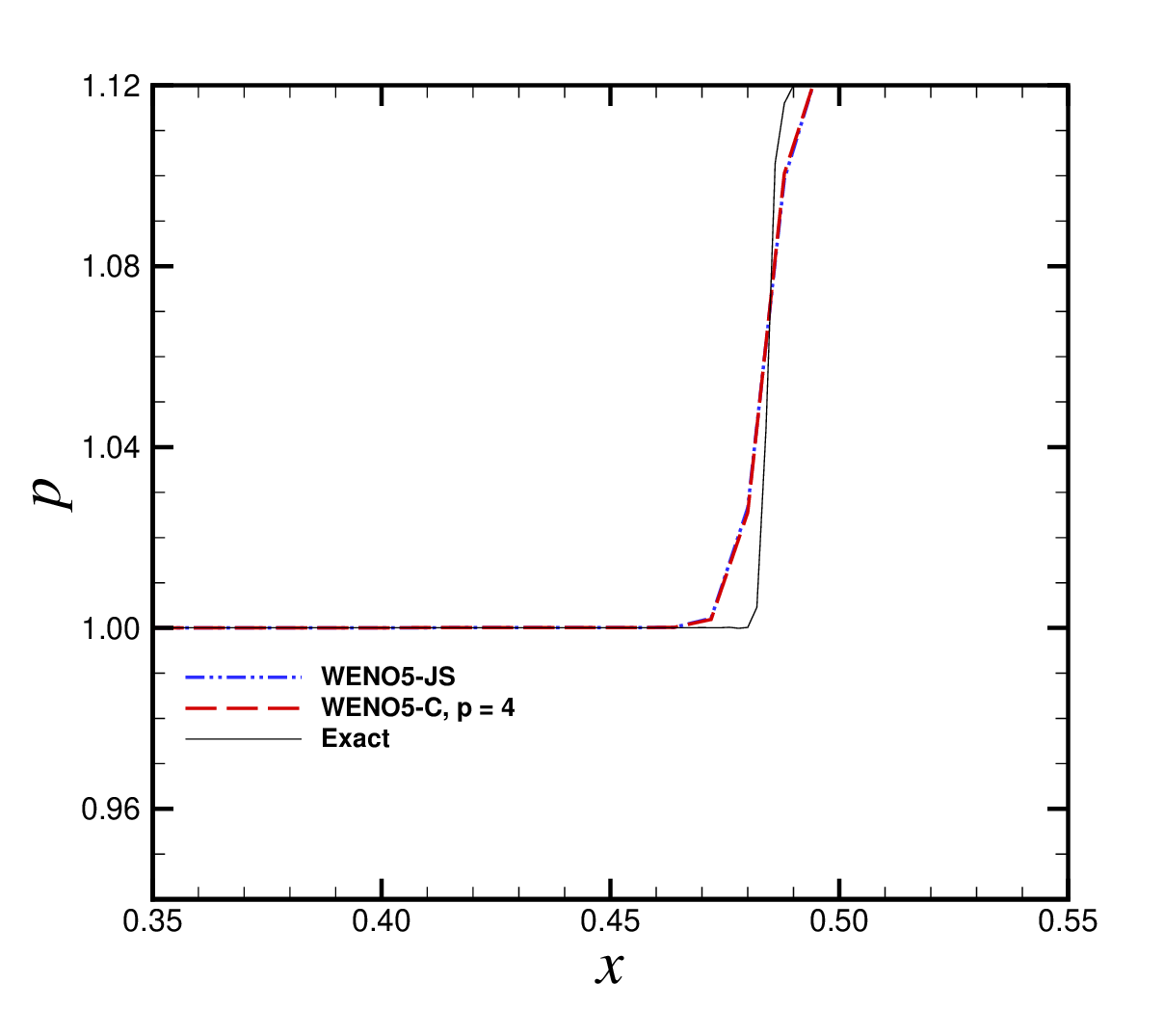}
	
	\includegraphics[width=0.46\textwidth]{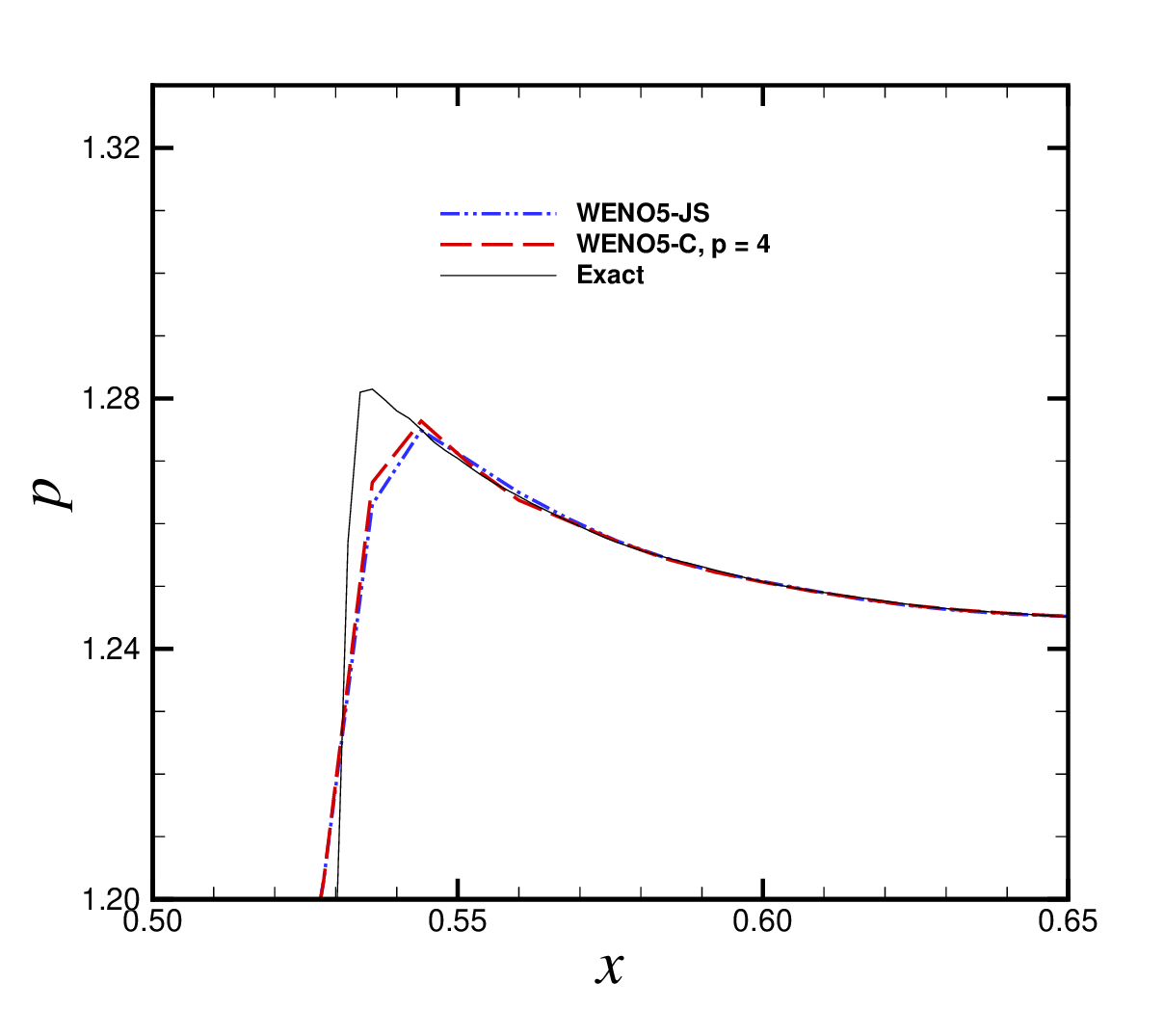}
	\includegraphics[width=0.46\textwidth]{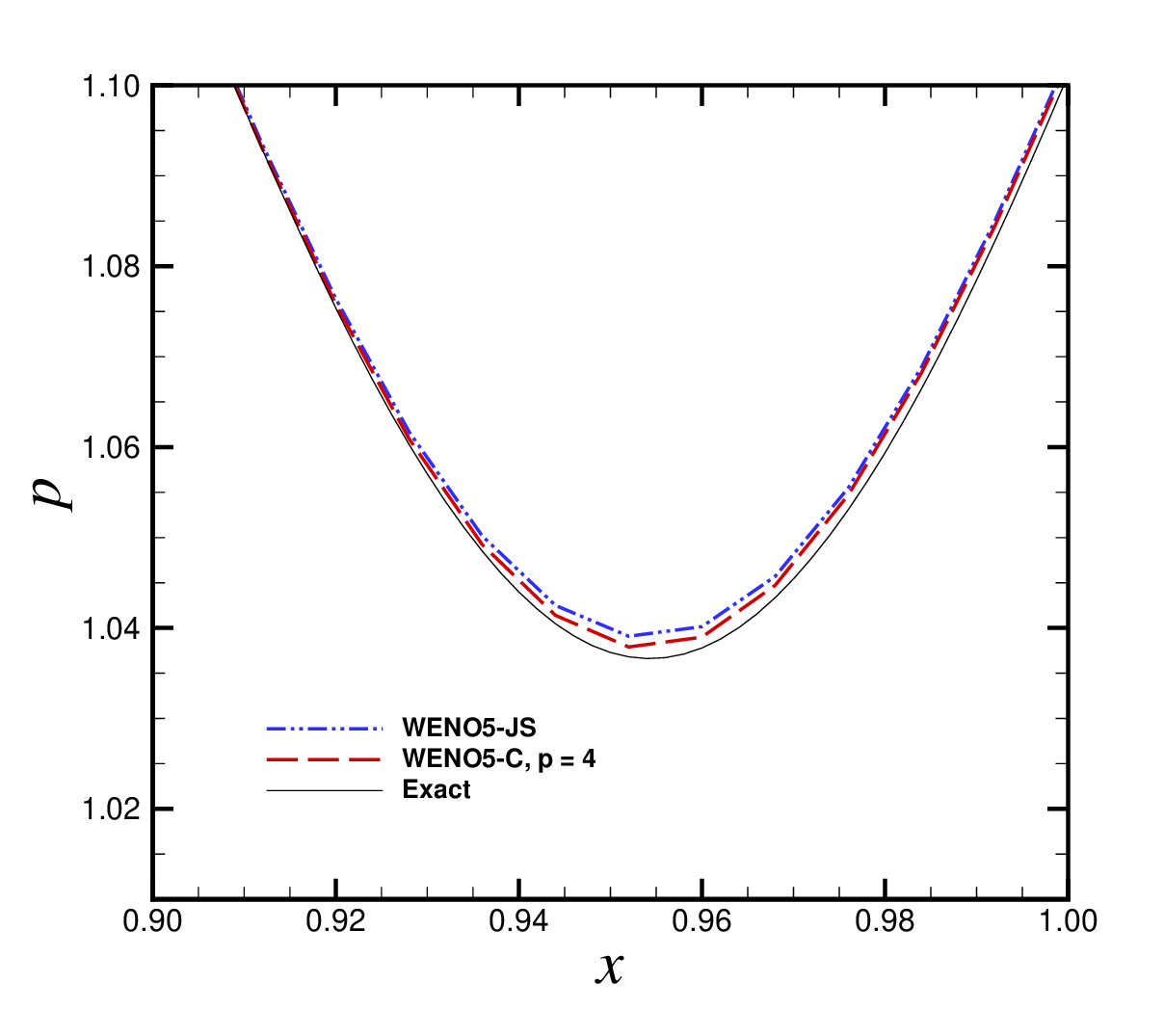}
	\caption{The pressure profile of the shock-vortex interaction problem along $y=0.5$ section at $0.6$.}
	\label{f:vortex} 
\end{figure} 

Figures~\ref{f:vortex} displays the pressure distribution along the $y=0.5$ section at the initial state and also at $t=0.6$ where the vortex has passed the shock. The results obtained by WENO5-JS and WENO5-C are compared immediately before and after the shock and also at the vortex core. At the vortex core and also after the shock, we observe more accurate results are obtained by the WENO5-C scheme. However, before the shock no difference is seen between the two schemes.

\section{Conclusions}
\label{conclusions}
We introduced a new method to improve the accuracy of the WENO schemes, which we called WENO-C. The ``C'' stands for combined. The goal was to use all the available sub-stencils of different orders, from $k$th- to $(2k-2)$th-order, to construct the flux. First, $(k-1)$ intermediate fluxes were constructed using a weighted combination of the sub-stencils of the same order of accuracy, similar to the conventional WENO schemes. Then, the intermediate fluxes were combined to form the final flux. The key idea was to define suitable weights (called total weights) for the linear combination of these intermediate fluxes. This was done by defining the  total weights as a function of their corresponding total smoothness indicator and a coefficient to control the contribution of each intermediate flux to the final flux. The total smoothness indicator of each intermediate flux was defined as a weighted combination of the smoothness indicators of its sub-stencils where each weight was the same as its corresponding flux weight in the construction of that intermediate flux. The numerical results showed that this way of defining the total smoothness indicators properly allows to use all the sub-stencils which does not intersect the discontinuity.

Furthermore, using a simple relation for the coefficients used in the total weights, we were able to increase the effect of higher-order sub-stencils and therefore increase the results accuracy. 
The numerical experiments showed that the new method while increased the overall accuracy of the results, was resistant against spurious oscillations near discontinuities and therefore suitably preserved the ENO property. 


\section*{Acknowledgments}
The authors would like to thank the University of Tehran for financial support for this research under grant number 04/1/28745.

\bibliographystyle{unsrt} 
\bibliography{mybibfile}

\end{document}